\documentclass[
aps,
prl,
reprint,
superscriptaddress
]{revtex4-2}

\usepackage{amsmath,amssymb,amsfonts,bm}
\usepackage{graphicx}
\usepackage{hyperref}
\usepackage{siunitx}
\usepackage[capitalise,noabbrev]{cleveref}
\usepackage{subcaption}
\usepackage{grffile}   
\usepackage{epstopdf}

\usepackage{caption}      
\usepackage{array,tabularx}
\usepackage{array}

\begin{document}

\title{Pairwise Distance-Diffusion Analysis (PDDA): A Geometric Framework for Estimating Hurst Exponents in Multivariate Long-Memory Processes}

\author{Diogo C.~Soriano}
\email{diogo.soriano@ufabc.edu.br}
\homepage{https://orcid.org/0000-0002-9804-7477}
\affiliation{Center for Engineering, Modeling, and Applied Social Sciences, Federal University of ABC (UFABC), São Bernardo do Campo, Brazil}

\author{Frederique Vanheusden}
\affiliation{School of Science and Technology, Nottingham Trent University, Nottingham, United Kingdom}

\author{Slawomir J.~Nasuto}
\affiliation{Biomedical Engineering, School of Biological Sciences, University of Reading, Reading, United Kingdom}

\begin{abstract}
We introduce Pairwise Distance--Diffusion Analysis (PDDA), a geometric
framework that connects the Hurst exponent to the scaling of pairwise
distances in long-memory stochastic processes. From a single distance-plot
representation, PDDA yields three complementary routes to persistence:
R/S--PDDA, based on the growth of geometric extrema; MSD--PDDA, based on
the scaling of the second moment of lagged distances; and recurrence-volume
scaling, in which the decay of close returns with increasing temporal
separation reflects the expansion of the lagged displacement cloud.
The framework extends naturally to multivariate isotropic and anisotropic
processes, where the local spatial dimension specifies the available degrees
of freedom while the Hurst exponents govern their temporal expansion.
These results establish the distance plot as a common geometric
representation of persistence, providing a unified distance-based foundation
for Hurst analysis.
\end{abstract}

\maketitle

\section{Introduction}

Long-range temporal dependence and anomalous diffusion are central features of complex systems across physics, biology, neuroscience, and economics. 
In biological and critical systems, long-range correlations are often discussed in terms of scale invariance and $1/f$ noise~\cite{Bak1987,Ferraz2025}, 
whereas in finance and econophysics phenomena are framed through fractal market hypotheses and persistent volatility~\cite{Peters1994, Gatheral2018Rough}. 
In parallel, anomalous diffusion theory focuses on single-particle dynamics, characterizing deviations of mean-squared displacement from classical Brownian scaling~\cite{Metzler2014,GomesFilho2025}. 
Despite differences in language and methodology, these perspectives share a common quantitative descriptor: the Hurst exponent $H$, which governs both long-memory effects and non-Brownian diffusive scaling~\cite{Mandelbrot1968Fractional,Beran1994,SamorodnitskyTaqqu1994,gao2007multiscale}.

Here we introduce \emph{Pairwise Distance--Diffusion Analysis} (PDDA),
a geometric framework that characterizes persistence from pairwise
distances between cumulative states of a stochastic process. For a
temporal separation $\tau$, the corresponding distance-plot diagonal
samples the lag-conditioned distribution $p(d|\tau)$. Different
Hurst-related observables emerge as functionals of this same object:
R/S--PDDA from the growth of distance extrema, and MSD--PDDA from its
second moment, connecting distance plots to the variance-scaling laws
of anomalous diffusion~\cite{Metzler2014,Sikora2017}. This extends the
distance--second-order-statistics correspondence of Rohde
\emph{et al.}~\cite{rohde2008stochastic} to long-memory scaling and
naturally generalizes to multivariate stochastic processes, investigated
here using fractionally integrated ARFIMA models with correlated
innovations.

A third route follows from recurrence-volume scaling. The decay of
close returns with temporal separation reflects the expansion of the
lagged displacement cloud: its local dimension specifies the available
spatial degrees of freedom, while the Hurst exponents govern their
temporal expansion. Since diagonal recurrence rates admit generalized
autocorrelation and Wiener--Khinchin-type spectral interpretations
~\cite{Marwan2007PhysRep,ZbilutMarwan2008}, this provides a direct bridge
between pairwise-distance geometry, temporal correlation, and spectral
scaling.

More broadly, PDDA identifies $p(d|\tau)$ as a common distance-based
representation of stochastic dynamics, suggesting a geometric framework
for multiscale analysis based on the diffusion of trajectory clouds in
state space. Such a formulation is potentially relevant wherever
persistence and scale-dependent fluctuations are central, including
anomalous diffusion, quantitative finance, and neuroscience. PDDA therefore shifts Hurst analysis from a collection of scaling estimators toward a common geometric description of multiscale stochastic dynamics.

\section{Theoretical Background}

\subsection{Reinterpreting the Hurst exponent from distance plots}
\label{sec:hurst_distanceplots}

The Hurst exponent, classically defined through rescaled–range (R/S) analysis, quantifies how fluctuations of a cumulative process scale with observation length~\cite{Peters1994, gao2007multiscale}.
Here we recast this scaling in geometric terms, expressing $H$ directly through pairwise distances between cumulative states and extending the formulation to higher-dimensional settings.

\medskip
\noindent\textbf{Route 1 (R/S-PDDA): Recasting R/S scaling via distance plots.}
\medskip
Let $\{X_t\}_{t=1}^N$ be a zero-mean stationary process and $Z_t$ the cumulative deviate path given its intergrated form:
\begin{equation}
Z_t = \sum_{i=1}^{t}(X_i - \bar X), \qquad  t = 1,\ldots,N.
\label{eq:cumpath}
\end{equation}

The Hurst exponent $H$ quantifies how the amplitude of its fluctuations grows with the time scale \cite{Peters1994, gao2007multiscale}. In the classical (R/S) analysis, the range and standard deviation over a window of length $n$ are defined as:
\begin{equation}
R(n) = \max_{1\le t \le n} Z_t - \min_{1\le t \le n} Z_t, \quad
S(n) = \sqrt{\frac{1}{n}\sum_{t=1}^{n}(X_t-\bar X)^2},
\label{eq:R/S_def}
\end{equation}
and the expected ratio scales as:
\begin{equation}
E\!\left[\frac{R(n)}{S(n)}\right] \propto n^H.
\label{eq:R/S_expect_ratio}
\end{equation}

For $H=0.5$, increments are uncorrelated (white noise, or Brownian motion); $H>0.5$ corresponds to persistent (or superdiffusion) behavior, while $H<0.5$ outlines anti-persistent (or subdiffusion) regimes \cite{Peters1994, gao2007multiscale, Metzler2014, GomesFilho2025}.
A distance-based formulation follows directly by defining the distance matrix of the cumulative trajectory \cite{Marwan2007PhysRep}:
\begin{equation}
D_{ij} = |Z_i - Z_j|, \qquad i,j=1,\ldots,N,
\label{eq:DP}
\end{equation}
and the cumulative-path diameter within a window of length $n$ can be obtained:
\begin{equation}
R_D(n) = \sqrt{\max_{i,j\le n}(Z_i-Z_j)^2} = \max_{i,j\le n} D_{ij}.
\label{eq:distance_range}
\end{equation}

If we consider $S_D(n)$ the standard deviation of the increments, this leads exactly to the classical R/S definition in the univariate scenario:   
\begin{equation}
E\!\left[\frac{R_D(n)}{S_D(n)}\right] \propto n^H.
\label{eq:DP_R/S_ratio}
\end{equation}

This formulation highlights that $H$ controls how the geometry of pairwise distances expands with observation length, recasting the Hurst exponent as a measure of the cumulative path’s geometric extent in distance–plot space. A purely geometric estimator ($\widehat{H}_{R/S}$) can then be constructed directly from $D_{ij}$.

\medskip
\noindent \textbf{Route 2 (MSD-PDDA): Obtaining $H$ from the distance profile and variance--scaling law.}
\medskip

For a fixed lag $\tau$, the entries
$D_{t,t+\tau}=|Z_{t+\tau}-Z_t|$ form the $\tau$-offset diagonal profile of the distance plot. The empirical time-averaged squared distance along this diagonal
\begin{equation}
\widehat{M}_2(\tau)
\equiv
\left\langle D_{t,t+\tau}^{\,2}\right\rangle_t
=
\frac{1}{N-\tau}
\sum_{t=1}^{N-\tau}
\left(Z_{t+\tau}-Z_t\right)^2,
\end{equation}
defines the empirical mean-squared displacement (MSD) between cumulative states separated by lag $\tau$. Its ensemble counterpart is
$M_2(\tau)=
\mathbb{E}\!\left[(Z_{t+\tau}-Z_t)^2\right]$.
Assuming finite second moments and centered displacements,
$\mathbb{E}[Z_{t+\tau}-Z_t]=0$, the expected squared distance satisfies
\begin{equation}
M_2(\tau)
=
\operatorname{Var}(Z_{t+\tau})
+
\operatorname{Var}(Z_t)
-
2\operatorname{Cov}(Z_{t+\tau},Z_t).
\end{equation}

This distance--covariance correspondence was established by Rohde \textit{et al.}~\cite{rohde2008stochastic}. Here, we explore its scaling implications for long-memory processes. For processes with stationary increments, $M_2(\tau)$ depends only on the lag and connects directly to the power-law variance growth underlying anomalous diffusion and fractional Brownian motion. 

Indeed, it is well-known that for long-memory increments the autocovariance of $X_t$ decays as \cite{gao2007multiscale}:
\begin{equation}
C_X(\tau) \sim \tau^{2H-2}, \qquad \tau \to \infty,
\label{eq:corr_decay}
\end{equation}

\noindent and the variance of the cumulative sum (the deviation process $Z_t$) obeys ~\cite{Beran1994,falconer2003fractal,gao2007multiscale}:
\begin{equation}
M_2(\tau) = \mathbb{E}\big[(Z_{t+\tau}-Z_t)^2\big]
\sim \tau^{2H},\qquad \tau\to\infty,
\label{eq:scaling_MSD}
\end{equation}

The empirical profile $\widehat{M}_2(\tau)$ estimates
$M_2(\tau)$ and, under stationary increments and the corresponding
ergodicity conditions, converges to the ensemble quantity as the
number of available displacement pairs increases, while inheriting its scaling law. 

For sufficiently long realizations, it is possible to estimate $\widehat{H}$ after taking the logarithms from both sides in Eq.~\eqref{eq:scaling_MSD} in the form:
\begin{equation}
\widehat{H}_{M_2} = \tfrac{1}{2}\,\frac{d(\log \widehat{M}_2(\tau))}{d(\log\tau)}.
\label{eq:Hurst_fromM2}
\end{equation}

The slope of $\log \widehat{M}_2(\tau)$ versus $\log\tau$ therefore \textit{provides a purely geometric estimator of the Hurst exponent}, derived directly from the distance plot without reference to ranges or standard deviations.
\begin{center}
\fbox{
\begin{minipage}{0.98\linewidth}
\noindent\textbf{Contribution 1 — Unified interpretation.}
PDDA unifies R/S and MSD scaling within a single distance-based formulation of the Hurst exponent, revealing both as complementary geometric projections of the same trajectory expansion law.
\end{minipage}
}
\end{center}

By operating in the distance–plot domain, this framework favors the generalization to multivariate systems, where pairwise distances replace the unidimensional notion of range, allowing the Hurst exponent to be defined in a coordinate–invariant geometric form.

\subsection*{Generalizing PDDA for multivariate long-memory processes}

To investigate the multivariate setting, we consider a parsimonious
fractionally integrated model representative of multivariate long-memory
processes. Let $\mathbf{X}_t \in \mathbb{R}^m$ be defined by:
\begin{equation}
X_t^{(k)}
=
\sum_{\ell \ge 0} a_\ell^{(k)}
\varepsilon_{t-\ell}^{(k)},
\qquad
\boldsymbol{\varepsilon}_t
\sim
\mathcal{N}(\mathbf{0},\Sigma_\rho),
\label{eq:multi_ARFIMA}
\end{equation}
where $\boldsymbol{\varepsilon}_t \sim
\mathcal{N}(\mathbf{0},\Sigma_\rho)$ denotes an $m$-variate
zero-mean Gaussian innovation vector with covariance matrix $\Sigma_\rho$, and each coordinate follows an ARFIMA$(0,d_k,0)$ fractional
filter~\cite{Hosking1981,Liu2017}, with:
\begin{equation}
a_0^{(k)}=1,
\qquad
a_\ell^{(k)}
=
a_{\ell-1}^{(k)}
\frac{\ell-1+d_k}{\ell},
\qquad
d_k=H_k-\frac{1}{2}.
\label{eq:arfimacoefs}
\end{equation}
This is the zero-AR case of multivariate fractionally integrated models \cite{SelaHurvich2009}, with coordinate-wise fractional filtering and cross-dependence set by the innovation covariance. For the bivariate case:
\begin{equation}
\Sigma_\rho =
\begin{pmatrix}
1 & \rho\\
\rho & 1
\end{pmatrix},
\label{eq:innovation_corr}
\end{equation}
where $\rho$ controls instantaneous innovation correlation.

This construction separates marginal temporal scaling from cross-coordinate dependence, unlike general multivariate fractional Brownian motion (fBm) models with additional cross-covariance and temporal-asymmetry parameters~\cite{Amblard2011mfBm}.
 
The multidimensional cumulative trajectory is defined as:
\begin{equation}
\mathbf{Z}_t = \sum_{i=1}^t (\mathbf{X}_i - \bar{\mathbf{X}}),
\label{eq:multi_Z}
\end{equation}
generalizing Eq.~(1), with the corresponding distance matrix:
\begin{equation}
D_{ij} = \lVert \mathbf{Z}_i - \mathbf{Z}_j \rVert ,
\label{eq:multi_D}
\end{equation}
in which $\lVert . \rVert$ denotes the Euclidean norm.

\medskip
\noindent\textbf{Route 1 (R/S-PDDA): DP-based rescaled–range geometry in $\mathbb{R}^m$}
\medskip

For a block of size $n$, a multivariate diameter and dispersion can be defined:
\begin{equation}
R_D(n) = \max_{i,j\le n}\lVert \mathbf{Z}_i - \mathbf{Z}_j\rVert,
\quad
S_D(n) = \sqrt{\operatorname{tr} \operatorname{Cov}(\mathbf{X}_1,\dots,\mathbf{X}_n)},
\label{eq:multi_RS}
\end{equation}

\noindent in which $\operatorname{tr} \operatorname{Cov}(.)$ is the trace of the covariance matrix. A related multivariate R/S construction based on the Euclidean
trajectory diameter was proposed in Ref.~\cite{Meraz2022}; here it is
recast directly in distance--plot space using a trace-based dispersion. This allows to obtain the analogue of the classical R/S law in the form:
\begin{equation}
\mathbb{E}\!\left[\frac{R_D(n)}{S_D(n)}\right]
\propto n^{H_{\mathrm{multi}}}.
\label{eq:multi_RS_scaling}
\end{equation}

\paragraph*{Isotropic case.}
If all coordinates share the same Hurst exponent 
$H_1=\dots=H_m = H$, then $H_{\mathrm{multi}} = H$. The multivariate trajectory is obtained through a fixed linear deformation of the isotropic process. Instantaneous cross-correlation modifies just geometric prefactors but preserves the Hurst asymptotic scaling exponent (see Supplementary Material S1.2 and S3.2).

\paragraph*{Anisotropic case.}
When $H_1,\dots,H_m$ differ, the diameter in~\eqref{eq:multi_RS_scaling}
is governed by the most persistent coordinate:
\begin{equation}
R_D(n)\;\asymp\; n^{H_{\max}}, 
\qquad 
H_{\max} := \max_k H_k.
\label{eq:max_track}
\end{equation}
Thus, $H_{\mathrm{multi}} = H_{\max}$, defining R/S-PDDA as max-persistence selector.

\medskip
\noindent\textbf{Route 2 (MSD-PDDA): Variance scaling via the distance profile}
\medskip

The distance--profile estimator extends naturally to the vector case:
\begin{equation}
\widehat{M}_2(\tau)
=
\left\langle
\left\|
\mathbf{Z}_{t+\tau}-\mathbf{Z}_t
\right\|^2
\right\rangle_t .
\end{equation}

Its ensemble counterpart is
$M_2(\tau)=
\mathbb{E}[
\|\Delta_\tau \mathbf{Z}_t\|^2
]$,
where
$\Delta_\tau \mathbf{Z}_t=\mathbf{Z}_{t+\tau}-\mathbf{Z}_t$.

Using the Euclidean norm, this quantity decomposes exactly into the coordinate-wise displacement contributions:
\begin{equation}
M_2(\tau)
=
\sum_{k=1}^{m}
\mathbb{E}
\left[
\left(
\Delta_\tau Z_t^{(k)}
\right)^2
\right].
\label{eq:M2coordinate}
\end{equation}

This identity does not require independence between coordinates.
For centered displacements, each term in Eq.~\eqref{eq:M2coordinate} corresponds to the
marginal displacement variance.

For fractional coordinates, each marginal displacement contribution
obeys its own variance--scaling law:
\begin{equation}
\mathbb{E}[(\Delta_\tau Z_t^{(k)})^2]
\sim c_k\tau^{2H_k}.
\label{eq:MarginalVarLaw} 
\end{equation}

Substituting Eq.~\eqref{eq:MarginalVarLaw} into Eq.~\eqref{eq:M2coordinate} yields:
\begin{equation}
M_2(\tau)
\sim
c_1\tau^{2H_1}
+
c_2\tau^{2H_2}
+
\cdots
+
c_m\tau^{2H_m},
\qquad c_k>0.
\label{EqMixedPowerLaw}
\end{equation}

\textit{Isotropic case.}
If all coordinates share the same Hurst exponent,
$H_k\equiv H$, Eq.~\eqref{EqMixedPowerLaw} reduces to:
\begin{equation}
M_2(\tau)
\sim
\left(
\sum_{k=1}^{m}c_k
\right)
\tau^{2H}
\propto
\tau^{2H},
\qquad
H_{\mathrm{multi}}=H.
\end{equation}

\paragraph*{Anisotropic case.} 
Eq.~\eqref{EqMixedPowerLaw} shows that $M_2(\tau)$ is a mixture of power laws.
Its slope in log--log coordinates,
\begin{equation}
H_{\mathrm{eff}}(\tau)
=
\frac{1}{2}
\frac{d\ln M_2(\tau)}
{d\ln\tau},
\end{equation}
is scale dependent:
\begin{equation}
H_{\mathrm{eff}}(\tau)
\approx
\frac{
\sum_{k=1}^{m}
H_k c_k \tau^{2H_k}
}{
\sum_{k=1}^{m}
c_k \tau^{2H_k}
}
\longrightarrow
H_{\max},
\qquad
\tau\to\infty.
\label{eq:Heff}
\end{equation}

Thus, within the mixed-scaling regime, $H_{\mathrm{eff}}(\tau)$ is a scale-dependent convex combination of the component exponents (details in Sec.~S1.1). At smaller lags within this regime, multiple components contribute according to
their relative weights, whereas at sufficiently large lags the terms
associated with $H_{\max}$ dominate.

\begin{center}
\fbox{
\begin{minipage}{0.98\linewidth}
\noindent\textbf{Contribution 2 — Multivariate generalization.}
We formulate the multivariate Hurst exponent geometrically through pairwise distances. R/S--PDDA tracks the dominant persistence direction, whereas MSD--PDDA captures scale-dependent mixtures of marginal exponents. Under the stated anisotropic scaling conditions,
both asymptotically approach $H_{\max}$.
\end{minipage}
}
\end{center}

These scaling relations show that pairwise distances encode how the cumulative trajectory expands across temporal scales. A complementary view is obtained by asking not how far the trajectory spreads, but how the lagged displacement cloud occupies local neighborhoods in state space. This leads naturally to recurrence-probability scaling.

\subsection{Recurrence-probability scaling and its relation to Hurst exponents}
\label{sec:rec_scalingLaw}

The distance plot also directly characterizes how temporal persistence
governs recurrence statistics. For a fixed lag $\tau$, consider the
probability that two cumulative states lie within a distance $\varepsilon$
of one another,
\begin{equation}
P(\varepsilon,\tau)
=
\Pr\!\left(
\|\mathbf Z_{t+\tau}-\mathbf Z_t\|\leq\varepsilon
\right).
\label{eq:RecProb}
\end{equation}

Geometrically, the lagged increments
$\Delta_\tau\mathbf Z_t=\mathbf Z_{t+\tau}-\mathbf Z_t$
form a displacement cloud whose characteristic extent along coordinate
$k$ scales as $\tau^{H_k}$. 
For anisotropic operator scaling, the lagged displacement cloud
expands under the dilation
$\tau^{\mathcal H}$, with
$\mathcal H=\mathrm{diag}(H_1,\ldots,H_m)$
~\cite{Didier2011}. Since volumes transform with the
determinant of the dilation operator~\cite{Bierme2007},
\begin{equation}
V(\tau)\propto\det(\tau^{\mathcal H})
=\prod_k\tau^{H_k}
=\tau^{\sum_k H_k}.
\end{equation}
 
A fixed small recurrence ball,
in contrast, has local volume proportional to $\varepsilon^m$ \cite{grassberger1983characterization,falconer2003fractal}.
Provided that the increment density is regular near the origin, the
fraction of the cloud contained in this ball consequently scales as
\begin{equation}
P(\varepsilon,\tau) \propto \frac{\varepsilon^m}{V(\tau)}
\propto
\underbrace{\varepsilon^m}_{\text{local geometry}}\,
\underbrace{\tau^{-\sum_{k=1}^{m}H_k}}_{\text{temporal expansion}} .
\label{eq:RecScaling}
\end{equation}

Thus, the local dimension $m$ specifies the available spatial degrees
of freedom, whereas the Hurst exponents determine how rapidly the
corresponding displacement volume expands with lag. In the isotropic
case, $H_k\equiv H$, Eq.~\eqref{eq:RecScaling} reduces to
$P(\varepsilon,\tau)\propto\varepsilon^m\tau^{-mH}$.

\begin{figure*}[t!]
\centering
\includegraphics[width=0.90\textwidth]{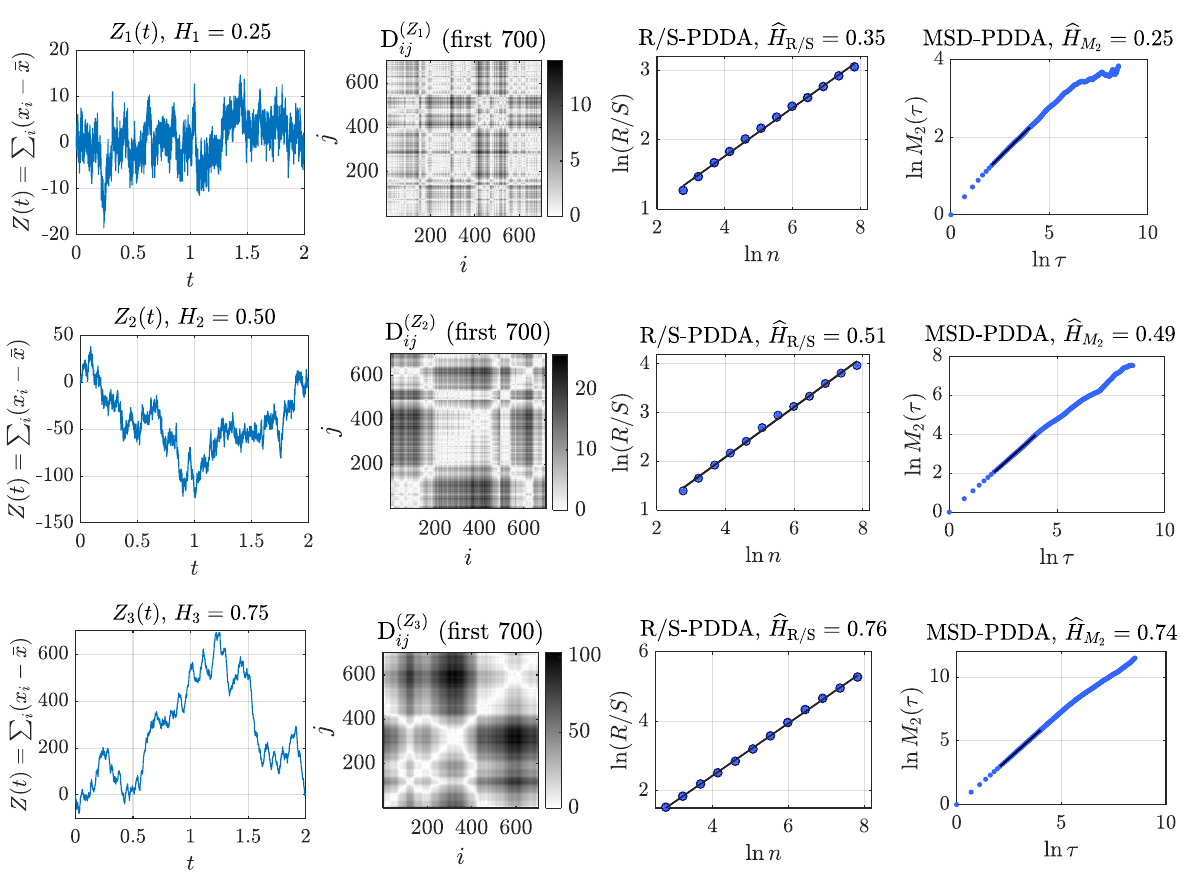}
\caption{\textbf{Univariate ARFIMA simulations and Hurst estimation via distance plots}
($N=2\times10^4$). Rows correspond to $H=0.25$, $0.50$, and $0.75$.
Columns show cumulative paths $Z(t)$, distance plots $D_{ij}^{(Z)}$,
R/S-PDDA scaling, and MSD-PDDA scaling. Increasing $H$ produces stronger
spatial coherence in the distance plots. Log--log slopes yield
$\widehat H_{R/S}$ and $\widehat H_{M_2}$ (panel titles).}
\label{fig:Exp1}
\end{figure*}

This scaling probes a different property from MSD--PDDA. Whereas
$M_2(\tau)$ asks how far the process typically moves and, under
anisotropy, progressively emphasizes the largest exponent,
$P(\varepsilon,\tau)$ asks what fraction of the displacement cloud
remains inside a fixed neighborhood and therefore probes its volume
expansion. In the bivariate case these complementary functionals
provide, asymptotically,
$\widehat H_{M_2}\rightarrow H_{\max}$ and
$\widehat\beta\rightarrow H_1+H_2$, where
$P(\varepsilon,\tau)\sim\tau^{-\beta}$. Hence,
$H_{\min}$ can be recovered from
$\widehat H_{\min}=\widehat\beta-\widehat H_{\max}$.

More generally, each $\tau$-diagonal of the distance plot provides an
empirical sample of the lag-conditioned distance distribution
$p(d|\tau)$. Accordingly, the diagonal average defining
$M_2(\tau)$ estimates its raw second moment,
$M_2(\tau)=\int_0^\infty d^2p(d|\tau)\,{\rm d}d$, whereas thresholding
the same distances gives the recurrence probability
$P(\varepsilon,\tau)=\int_0^\varepsilon p(d|\tau)\,{\rm d}d$. The former corresponds to the diagonal-wise recurrence rate
$RR_\tau(\varepsilon)$, interpreted in RQA as a generalized
autocorrelation measure~\cite{Marwan2007PhysRep}. Recurrence-based
extensions of the Wiener--Khinchin relation further connect
$RR_\tau$ to generalized spectral representations~\cite{ZbilutMarwan2008},
providing a bridge from pairwise-distance geometry to the
time--frequency duality underlying long-memory analysis \cite{gao2007multiscale}.
Despite these connections, the scaling relation between recurrence
decay and the Hurst exponent is not generally one-to-one.
Equation~\eqref{eq:RecScaling} makes this relation explicit:
$RR_\tau(\varepsilon)\sim\tau^{-mH}$ isotropically, whereas under
anisotropy its decay exponent becomes $\sum_k H_k$.
Thus, recurrence scaling probes the temporal expansion of the
multidimensional displacement volume rather than $H$ alone.

\begin{center}
\fbox{
\begin{minipage}{0.98\linewidth}
\textbf{Contribution 3 --- Recurrence-volume scaling as a third route to Hurst exponents.}
The temporal expansion of the lagged displacement cloud causes
recurrences within a fixed radius to decay with scale.
This yields $P(\varepsilon,\tau)\sim\tau^{-mH}$ for isotropic
processes and $\sim\tau^{-\sum_k H_k}$ under anisotropy,
providing a distinct geometric estimator of persistence from
$p(d|\tau)$ and linking Hurst scaling to generalized
autocorrelation decay.
\end{minipage}
}
\end{center}

The same pairwise-distance formulation also connects temporal
persistence to spatial occupancy. Whereas fixed-lag recurrence probes
the local geometry of the displacement cloud, the small-scale geometry
of pairwise separations over the full trajectory relates to its range
dimension, $D_{\mathrm{range}}=\min\{m,1/H\}$ for isotropic
fBm~\cite{adler1981geometry,kahane1985some, WuXiao2006}.
Accessing this global limit poses the familiar finite-sample challenges
of correlation-integral methods~\cite{grassberger1983characterization}.
Under anisotropy, it additionally depends on the full Hurst spectrum ~\cite{Sonmez2018} and lies beyond the present scope.

\section{Results}

In the following, the theoretical framework is validated using ARFIMA simulations across a broad range of parameters. Details concerning the simulations can be found in the supplemental material (S2).

\paragraph*{Univariate Scenario - Equivalence.} Figure~\ref{fig:Exp1} validates the equivalence between classical R/S analysis and its distance-based counterpart.
For ARFIMA processes spanning anti-persistent, Brownian, and persistent regimes ($H=0.25,0.5,0.75$), both R/S--PDDA and MSD--PDDA recover the theoretical scaling $\ln M_2(\tau)\sim 2H\ln\tau$.
While both estimators are asymptotically consistent, MSD--PDDA exhibits reduced bias in the anti-persistent regime.

Estimator performance is summarized in Tables~\ref{tab:comparison_smallN} and~\ref{tab:comparison_largeN}.
For intermediate sample sizes, R/S--PDDA overestimates $H$ in the anti-persistent regime ($H<0.5$), whereas MSD--PDDA achieves lower RMSE; for strongly persistent dynamics ($H\gtrsim0.65$), R/S--PDDA becomes slightly more accurate.
In the asymptotic regime (Table~\ref{tab:comparison_largeN}), both estimators converge, with MSD--PDDA retaining lower variance across all $H$ values.


\begin{table}[ht]
\centering
\caption{\textbf{Estimator comparison for $N=2048$ (intermediate–sample regime).}}
\label{tab:comparison_smallN}

\footnotesize
\setlength{\tabcolsep}{3pt}
\begin{tabular}{lccc|ccc}
\toprule
& \multicolumn{3}{c|}{\textbf{R/S-PDDA}} & \multicolumn{3}{c}{\textbf{MSD-PDDA}}\\
\textbf{$H_T$} & Bias & SD & RMSE & Bias & SD & RMSE\\
0.10 & 0.1699 & 0.0236 & 0.1715 & 0.0532 & 0.0372 & \textbf{0.0649}\\
0.25 & 0.1075 & 0.0284 & 0.1112 & 0.0170 & 0.0393 & \textbf{0.0426}\\
0.35 & 0.0698 & 0.0294 & 0.0757 & $-0.0022$ & 0.0387 & \textbf{0.0385}\\
0.50 & 0.0379 & 0.0351 & 0.0516 & $-0.0113$ & 0.0454 & \textbf{0.0465}\\
0.65 & 0.0107 & 0.0408 & \textbf{0.0420} & $-0.0189$ & 0.0458 & 0.0493\\
0.75 & $-0.0044$ & 0.0480 & \textbf{0.0480} & $-0.0401$ & 0.0478 & 0.0622\\
0.90 & $-0.0601$ & 0.0470 & \textbf{0.0761} & $-0.0912$ & 0.0460 & 0.1021\\
\end{tabular}
\end{table}

\begin{table}[ht]
\centering
\caption{\textbf{Estimator comparison for $N=32768$ (high–sample regime).}}
\label{tab:comparison_largeN}
\setlength{\tabcolsep}{3pt}
\begin{tabular}{lccc|ccc}
\toprule
& \multicolumn{3}{c|}{\textbf{R/S-PDDA}} & \multicolumn{3}{c}{\textbf{MSD-PDDA}}\\
\textbf{$H_T$} & Bias & SD & RMSE & Bias & SD & RMSE\\
0.10 & 0.1302 & 0.0078 & 0.1304 & 0.0582 & 0.0076 & \textbf{0.0587}\\
0.25 & 0.0733 & 0.0102 & 0.0740 & 0.0188 & 0.0097 & \textbf{0.0211}\\
0.35 & 0.0491 & 0.0122 & 0.0505 & 0.0043 & 0.0103 & \textbf{0.0111}\\
0.50 & 0.0222 & 0.0161 & 0.0274 & $-0.0009$ & 0.0121 & \textbf{0.0121}\\
0.65 & 0.0070 & 0.0170 & 0.0183 & $-0.0031$ & 0.0119 & \textbf{0.0122}\\
0.75 & $-0.0036$ & 0.0200 & 0.0202 & $-0.0054$ & 0.0113 & \textbf{0.0125}\\
0.90 & $-0.0353$ & 0.0228 & 0.0420 & $-0.0329$ & 0.0161 & \textbf{0.0366}\\
\end{tabular}
\end{table}

\begin{figure}[t]
\centering
\includegraphics[width=\columnwidth]{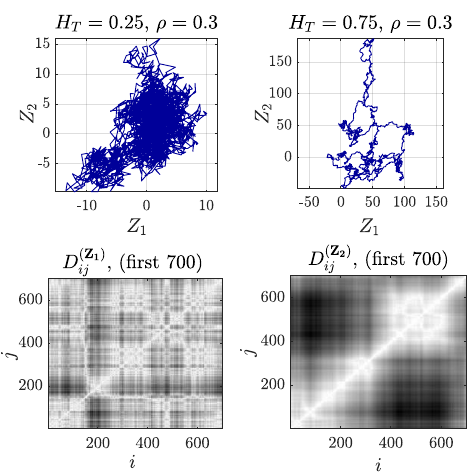}
\caption{\textbf{Bivariate isotropic ARFIMA: anti- vs persistent regimes}
($N=3000$, $\rho=0.3$).
Columns correspond to isotropic processes with $H_1=H_2=0.25$ (left) and $H_1=H_2=0.75$ (right). Rows show the cumulative trajectory $\mathbf Z$ (top) and the corresponding distance plot $D^{(\mathbf Z)}_{ij}$ computed from the first 700 samples (bottom). For the same realizations, the estimated exponents were (left)
$\widehat H_{R/S}=0.374$ and $\widehat H_{M_2}=0.267$ for $H_T=0.25$, and (right) $\widehat H_{R/S}=0.721$ and $\widehat H_{M_2}=0.712$ for $H_T=0.75$.}
\label{fig:2D_anti_vs_persistent}
\end{figure}

\begin{figure}[t]
\centering
\includegraphics[width=\linewidth]{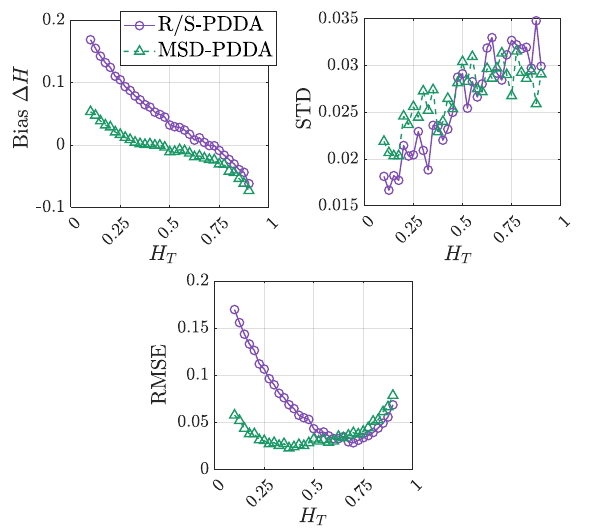}
\caption{\textbf{Estimator behavior across Hurst regimes in bivariate isotropic ARFIMA.}
Results for $\rho=0.3$, $N=3000$, and $R=100$. Panels show bias (left), standard deviation (center), and RMSE (right) of $\widehat H_{R/S}$ (purple) and $\widehat H_{M_2}$ (green) versus $H_T$.}
\label{fig:H_sweep_rho0p3}
\end{figure}

\paragraph*{Multivariate isotropic and anisotropic processes.}
Figures~\ref{fig:2D_anti_vs_persistent} and~\ref{fig:H_sweep_rho0p3} characterize the behavior of the distance-based estimators under isotropic multivariate dynamics ($H_1=H_2$).
Figure~\ref{fig:2D_anti_vs_persistent} shows that, independently of the dimensionality of the process, distance plots provide an effective two-dimensional representation of the cumulative trajectory, whose texture directly reflects the spatial coherence associated with the Hurst exponent. In this setting, MSD-PDDA consistently yields more accurate estimates in the anti-persistent regime, whereas both routes perform comparably for persistent dynamics at this sample size.

Figure~\ref{fig:H_sweep_rho0p3} quantifies these trends across $H_T$. Both estimators exhibit consistent scaling over the full range of persistence, confirming the robustness of the geometric formulation. MSD-PDDA achieves minimal bias and lowest RMSE around $H_T\approx0.45$ and remains superior throughout the anti-persistent and weakly persistent regimes, while R/S-PDDA becomes preferable only for strongly persistent processes ($H_T\gtrsim0.65$). These results show that both estimators probe complementary aspects of persistence: MSD-PDDA provides a statistically efficient route for low-sample and anti-persistent dynamics, whereas R/S-PDDA becomes competitive as long-range coherence dominates.

\begin{figure*}[t]
\centering
\begin{subfigure}[t]{0.48\textwidth}
\centering
\includegraphics[
height=5.4cm,
keepaspectratio
]{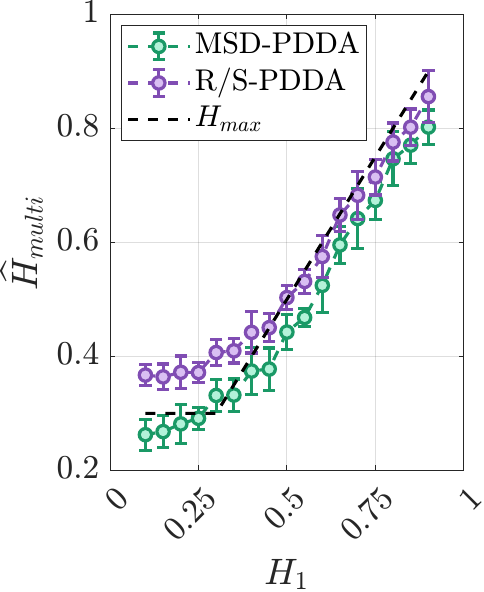}
\caption{Anisotropic sweep of $H_1$ with $H_2=0.30$}
\end{subfigure}
\hfill
\begin{subfigure}[t]{0.48\textwidth}
\centering
\includegraphics[
height=5.4cm,
keepaspectratio
]{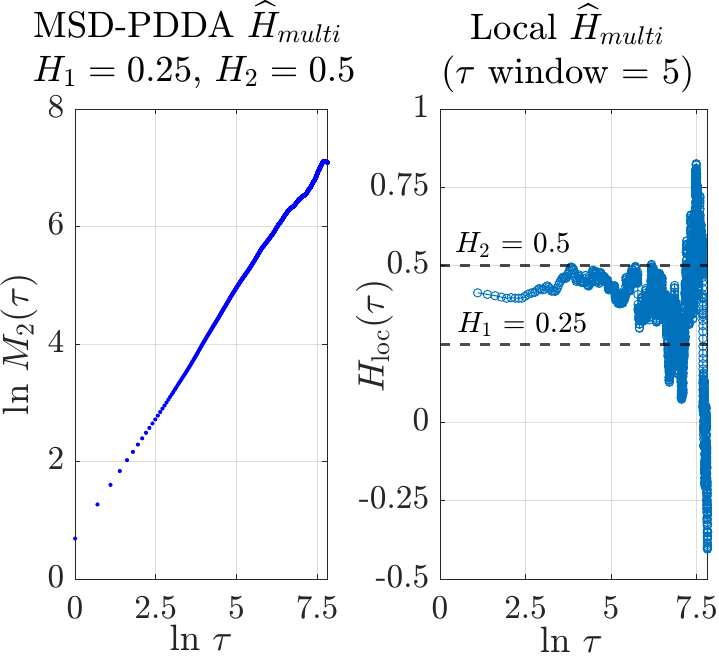}
\caption{Local MSD-PDDA scaling for $(H_1,H_2)=(0.25,0.50)$}
\end{subfigure}
\caption{\textbf{Bivariate anisotropic ARFIMA: R/S- vs MSD-PDDA.}
(a) Estimated multivariate exponent $\widehat H_{\mathrm{multi}}$ versus $H_1$ for fixed $H_2=0.30$ and $\rho=0.3$ ($N=3000$, $R=10$). Symbols show mean $\widehat H_{\mathrm{multi}}$ with $\pm$1 SD; dashed line denotes $H_{\max}=\max\{H_1,H_2\}$. 
(b) MSD-PDDA scaling for $(H_1,H_2)=(0.25,0.50)$: $\ln M_2(\tau)$ vs.\ $\ln\tau$ (left) and local slope $H_{\mathrm{loc}}(\tau)$ (right), showing the transition from mixed to dominant scaling.}
\label{fig:anisotropic_Hmulti}
\end{figure*}

In the anisotropic setting, we simulte ARFIMA processes with fixed $H_2=0.30$ and $\rho=0.3$, and sweep $H_1\in[0.10,0.90]$. $H_{\max}=\max\{H_1,H_2\}$ was considered as the reference exponent. Figure~\ref{fig:anisotropic_Hmulti}(a) shows that R/S-PDDA tracks $H_{\max}$ once $H_1>H_2$, but overestimates persistence in the strongly anti-persistent regime ($H_1<H_2$), consistent with diameter-based dominance by the most persistent direction. MSD-PDDA follows the overall trend but systematically underestimates $H_{\max}$.

Figure~4(b) explains this behavior for a representative case
$(H_1,H_2)=(0.25,0.50)$. The empirical local slope
$H_{\mathrm{loc}}(\tau)$ estimates the population quantity
$H_{\mathrm{eff}}(\tau)$ and reflects a scale-dependent mixture of the component exponents over the fitted scaling range. At larger
lags, it increasingly approaches the dominant exponent, consistently
with Eq.~\eqref{eq:Heff}. Since MSD--PDDA uses an intermediate fitting window, the resulting estimate generally lies between $H_{\min}$ and $H_{\max}$, with its precise value determined by the relative
coordinate weights within that window. This accounts for the
controlled underestimation of $H_{\max}$ observed in panel~(a).

\paragraph*{Recurrence-volume scaling.}
Figure~5 tests $P(\varepsilon,\tau)\sim\tau^{-\beta}$ at fixed
$\varepsilon=0.2$. For $H_1=H_2=0.25$,
$\widehat{\beta}=0.569$ approaches the isotropic prediction
$mH=0.50$ ($\widehat H=0.285$). Under anisotropy
($H_1=0.45$, $H_2=0.75$), $\widehat{\beta}=1.139$ approaches
$H_1+H_2=1.20$; combined with the large-scale MSD--PDDA estimate
$\widehat H_{\max}=0.714$, this yields
$\widehat H_{\min}=0.425$ ($H_{T} = 0.45$).
Thus, the two distance-based functionals recover complementary
information on the anisotropic Hurst spectrum.


\begin{figure}[t]
\centering
\includegraphics[width=\linewidth]{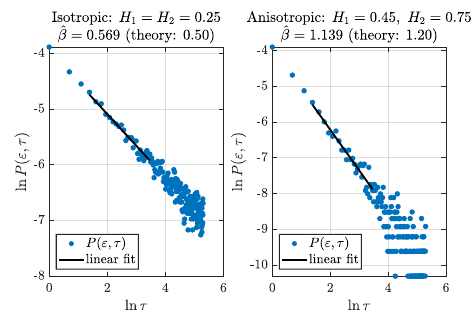}
\caption{\textbf{Recurrence-volume scaling in bivariate fractional processes.}
Recurrence probability $P(\varepsilon,\tau)$ at fixed $\varepsilon=0.2$
for isotropic ($H_1=H_2=0.25$; left) and anisotropic
($H_1=0.45$, $H_2=0.75$; right) processes.
Log--log fits yield $\widehat{\beta}=0.569$ and $1.139$, respectively,
consistent with the predicted scaling
$P(\varepsilon,\tau)\sim\tau^{-\beta}$, where
$\beta=mH$ under isotropy and $\beta=\sum_k H_k$ under anisotropy.}
\label{fig:RecScaling}
\end{figure}


\section{Discussion}

PDDA establishes the distance plot as a common geometric representation of long-memory stochastic processes. Its diagonal distance profiles sample the lag-conditioned distribution of pairwise distances, $p(d|\tau)$, which emerges as the natural object underlying this framework. Within this representation, R/S--PDDA, MSD--PDDA, lag-dependent recurrence measures (e.g., $RR_\tau$), become complementary statistical functionals of the same distribution. The Hurst exponent therefore governs both the trajectory scaling and the evolution of pairwise-distance statistics across temporal scales. Methods that have traditionally evolved independently - including long-range dependence analysis, anomalous diffusion, classical Hurst analysis, and RQA - are thus connected through a common pairwise-distance geometry, providing a foundation for geometric descriptors of stochastic dynamics beyond second-order statistics.

Because PDDA depends only on pairwise distances, it naturally extends to high-dimensional and multivariate stochastic systems. This formulation is particularly relevant where persistence and fractional dynamics play a central role, including fractional diffusion models for generative artificial intelligence \cite{Song2021SDE,Ho2020DDPM,Nobis2024GFDM}, rough volatility in quantitative finance \cite{Gatheral2018Rough,Kristoufek2014Efficiency}, and long-range temporal correlations in neuroscience and medical imaging \cite{Linkenkaer2001,Maxim2005}. By encoding persistence through pairwise-distance statistics, PDDA thus provides a geometric route for characterizing persistence in multivariate stochastic dynamics.


\section{acknowledgments}
DCS acknowledges support from CNPq (31397/2023-8) and FAPESP (2019/09512-0).

\section{Author Contributions}

D.C.S.: Conceptualization, Formal analysis, Software, Writing -  original draft. F.V.: Conceptualization, Writing - review and editing. S.J.N.: Conceptualization, Writing - review and editing.

\section{Data Availability}

All MATLAB codes required to reproduce the figures and numerical results reported in this work are publicly available at

\begin{center}
\url{https://github.com/dcsoriano/GitHub---PDDA}
\end{center}


\bibliographystyle{apsrev4-2}
\bibliography{refs}

@book{Beran1994,
  author    = {Beran, Jan},
  title     = {Statistics for Long-Memory Processes},
  publisher = {Chapman and Hall},
  year      = {1994},
  address   = {New York},
  series    = {Monographs on Statistics and Applied Probability},
  volume    = {61}
}

@book{SamorodnitskyTaqqu1994,
  author    = {Samorodnitsky, Gennady and Taqqu, Murad S.},
  title     = {Stable Non-Gaussian Random Processes: Stochastic Models with Infinite Variance},
  year      = {1994},
  publisher = {Chapman and Hall},
  address   = {New York}
}

@article{Mandelbrot1968Fractional,
  author  = {Mandelbrot, Benoit B. and Van Ness, John W.},
  title   = {Fractional Brownian Motions, Fractional Noises and Applications},
  journal = {SIAM Review},
  volume  = {10},
  number  = {4},
  pages   = {422--437},
  year    = {1968},
  url     = {https://www.jstor.org/stable/2027184},
  issn    = {0036-1445},
  note    = {Stable URL: https://www.jstor.org/stable/2027184}
}

@article{Marwan2007PhysRep,
  title        = {Recurrence plots for the analysis of complex systems},
  author       = {Marwan, Norbert and Romano, M. Carmen and Thiel, Marco and Kurths, J{\"u}rgen},
  journal      = {Physics Reports},
  volume       = {438},
  number       = {5-6},
  pages        = {237--329},
  year         = {2007},
  issn         = {0370-1573},
  doi          = {10.1016/j.physrep.2006.11.001},
  url          = {https://www.sciencedirect.com/science/article/abs/pii/S0370157306004066}
}

@book{gao2007multiscale,
  title     = {Multiscale Analysis of Complex Time Series: Integration of Chaos and Random Fractal Theory, and Beyond},
  author    = {Gao, Jianbo and Cao, Yinhe and Tung, Wen-wen and Hu, Jing},
  year      = {2007},
  publisher = {John Wiley \& Sons},
  address   = {Hoboken, NJ},
  isbn      = {9780471654704},
  doi       = {10.1002/9780470191651},
  url       = {https://doi.org/10.1002/9780470191651}
}

@article{rohde2008stochastic,
  title   = {Stochastic analysis of recurrence plots with applications to the detection of deterministic signals},
  author  = {Rohde, Gustavo K. and Nichols, Jonathan M. and Dissinger, Bryan M. and Bucholtz, Frank},
  journal = {Physica D: Nonlinear Phenomena},
  volume  = {237},
  number  = {5},
  pages   = {619--629},
  year    = {2008},
  issn    = {0167-2789},
  doi     = {10.1016/j.physd.2007.10.008},
  publisher = {Elsevier}
}

@book{falconer2003fractal,
  title     = {Fractal Geometry: Mathematical Foundations and Applications},
  author    = {Falconer, Kenneth},
  year      = {2003},
  publisher = {John Wiley \& Sons},
  address   = {Chichester},
  isbn      = {9780470848623},
}

@article{Metzler2014,
  author  = {Metzler, Ralf and Jeon, Jae-Hyung and Cherstvy, Andrey G. and Barkai, Eli},
  title   = {Anomalous diffusion models and their properties: non-stationarity, non-ergodicity, and ageing at the centenary of single particle tracking},
  journal = {Physical Chemistry Chemical Physics},
  year    = {2014},
  volume  = {16},
  number  = {38},
  pages   = {24128--24164},
  doi     = {10.1039/C4CP03465A},
  url     = {https://doi.org/10.1039/C4CP03465A}
}

@article{Sikora2017,
  author  = {Sikora, Grzegorz and Burnecki, Krzysztof and Wy{\l}oma{\'n}ska, Agnieszka},
  title   = {Mean-squared-displacement statistical test for fractional {B}rownian motion},
  journal = {Physical Review E},
  year    = {2017},
  volume  = {95},
  number  = {3},
  pages   = {032110},
  doi     = {10.1103/PhysRevE.95.032110},
  url     = {https://doi.org/10.1103/PhysRevE.95.032110}
}

@article{WuXiao2006,
  author  = {Wu, Dongsheng and Xiao, Yimin},
  title   = {Dimensional Properties of Fractional {B}rownian Motion},
  journal = {Journal of Theoretical Probability},
  year    = {2006},
  volume  = {20},
  number  = {2},
  pages   = {203--228},
  doi     = {10.1007/s10959-007-0075-8},
  url     = {https://doi.org/10.1007/s10959-007-0075-8}
}

@article{GomesFilho2025,
  author  = {Gomes-Filho, M{\'{a}}rcio Sampaio and Lapas, Luciano Calheiros and Gudowska-Nowak, Ewa and Oliveira, Fernando Albuquerque},
  title   = {The fluctuation-dissipation relations: {G}rowth, diffusion, and beyond},
  journal = {Physics Reports},
  year    = {2025},
  volume  = {1141},
  pages   = {1--43},
  doi     = {10.1016/j.physrep.2025.07.003},
  url     = {https://doi.org/10.1016/j.physrep.2025.07.003},
}

@article{Ferraz2025,
  author  = {Ferraz, Mariana Sacrini Ayres and Muotri, Alysson R. and Kihara, Alexandre Hiroaki},
  title   = {Deciphering Complexity in Human Brain Organoids via a Novel Hurst Exponent Estimation},
  journal = {Physical Review Letters},
  year    = {2025},
  volume  = {135},
  number  = {10},
  pages   = {108402},
  month   = {Sep},
  doi     = {10.1103/kl6z-ctdd},
  url     = {https://doi.org/10.1103/kl6z-ctdd}
}

@book{Peters1994,
  author    = {Peters, Edgar E.},
  title     = {Fractal Market Analysis: Applying Chaos Theory to Investment and Economics},
  publisher = {John Wiley \& Sons},
  year      = {1994},
  series    = {Wiley Finance Editions},
  address   = {New York},
  isbn      = {978-0471585244},
}

@article{Bak1987,
  title = {Self-organized criticality: An explanation of the 1/f noise},
  author = {Bak, Per and Tang, Chao and Wiesenfeld, Kurt},
  journal = {Physical Review Letters},
  volume = {59},
  issue = {4},
  pages = {381--384},
  year = {1987},
  doi = {10.1103/PhysRevLett.59.381}
}

@article{Hosking1981,
  author  = {Hosking, J. R. M.},
  title   = {Fractional Differencing},
  journal = {Biometrika},
  year    = {1981},
  volume  = {68},
  number  = {1},
  pages   = {165--176},
  doi     = {10.1093/biomet/68.1.165},
  url     = {https://doi.org/10.1093/biomet/68.1.165}
}

@article{Liu2017,
  author  = {Liu, Kai and Chen, YangQuan and Zhang, Xi},
  title   = {An Evaluation of {ARFIMA} ({Autoregressive Fractional Integral Moving Average}) Programs},
  journal = {Axioms},
  year    = {2017},
  volume  = {6},
  number  = {2},
  pages   = {16},
  doi     = {10.3390/axioms6020016},
  url     = {https://doi.org/10.3390/axioms6020016}
}

@article{grassberger1983characterization,
  title={Characterization of strange attractors},
  author={Grassberger, Peter and Procaccia, Itamar},
  journal={Physical Review Letters},
  volume={50},
  number={5},
  pages={346--349},
  year={1983},
  publisher={APS}
}

@article{ZbilutMarwan2008,
  author  = {Zbilut, Joseph P. and Marwan, Norbert},
  title   = {The {Wiener--Khinchin} theorem and recurrence quantification},
  journal = {Physics Letters A},
  volume  = {372},
  pages   = {6622--6626},
  year    = {2008},
  doi     = {10.1016/j.physleta.2008.09.027}
}

@book{adler1981geometry,
  title     = {The Geometry of Random Fields},
  author    = {Adler, Robert J.},
  year      = {1981},
  publisher = {John Wiley \& Sons},
  address   = {Chichester},
  isbn      = {0-471-27844-0}
}

@book{kahane1985some,
  title     = {Some Random Series of Functions},
  author    = {Kahane, Jean-Pierre},
  year      = {1985},
  edition   = {Second},
  publisher = {Cambridge University Press},
  series    = {Cambridge Studies in Advanced Mathematics},
  volume    = {5},
  address   = {Cambridge},
  isbn      = {0-521-45602-9}
}

@article{Linkenkaer2001,
  author  = {Klaus Linkenkaer-Hansen and Vadim V. Nikouline and J. Matias Palva and Risto J. Ilmoniemi},
  title   = {Long-Range Temporal Correlations and Scaling Behavior in Human Brain Oscillations},
  journal = {Journal of Neuroscience},
  volume  = {21},
  number  = {4},
  pages   = {1370--1377},
  year    = {2001},
  doi     = {10.1523/JNEUROSCI.21-04-01370.2001}
}

@article{Gatheral2018Rough,
  author  = {Jim Gatheral and Thibault Jaisson and Mathieu Rosenbaum},
  title   = {Volatility is Rough},
  journal = {Quantitative Finance},
  volume  = {18},
  number  = {6},
  pages   = {933--949},
  year    = {2018},
  doi     = {10.1080/14697688.2017.1393551}
}

@article{Kristoufek2014Efficiency,
  author  = {Ladislav Kristoufek and Miloslav Vosvrda},
  title   = {Measuring Capital Market Efficiency: Global and Local Correlation Structure},
  journal = {Physica A: Statistical Mechanics and its Applications},
  volume  = {392},
  number  = {1},
  pages   = {184--193},
  year    = {2013},
  doi     = {10.1016/j.physa.2012.08.003}
}

@inproceedings{Nobis2024GFDM,
  author = {Gabriel Nobis and Maximilian Springenberg and Marco Aversa and
            Rembert Daems and Roderick Murray-Smith and Sebastian Lapuschkin and
            Stefano Ermon and Manfred Opper and Christoph Knochenhauer and
            Michael Detzel and Shinichi Nakajima and Tolga Birdal and
            Luis Oala and Wojciech Samek},
  title = {Generative Fractional Diffusion Models},
  booktitle = {Advances in Neural Information Processing Systems},
  year = {2024}
}

@inproceedings{Ho2020DDPM,
  author    = {Jonathan Ho and Ajay N. Jain and Pieter Abbeel},
  title     = {Denoising Diffusion Probabilistic Models},
  booktitle = {Advances in Neural Information Processing Systems},
  volume    = {33},
  pages     = {6840--6851},
  year      = {2020},
  publisher = {Curran Associates, Inc.},
  url       = {https://proceedings.neurips.cc/paper/2020/hash/4c5bcfec8584af0d967f1ab10179ca4b-Abstract.html}
}

@article{Maxim2005,
  author  = {Voichiţa Maxim and Levent Şendur and Jalal Fadili and
             John Suckling and Rebecca Gould and
             Rob Howard and Ed Bullmore},
  title   = {Fractional Gaussian Noise, Functional MRI and Alzheimer's Disease},
  journal = {NeuroImage},
  volume  = {25},
  number  = {1},
  pages   = {141--158},
  year    = {2005},
  month   = mar,
  doi     = {10.1016/j.neuroimage.2004.10.044}
}

@inproceedings{Song2021SDE,
  author = {Yang Song and Jascha Sohl-Dickstein and Diederik P. Kingma and
            Abhishek Kumar and Stefano Ermon and Ben Poole},
  title = {Score-Based Generative Modeling through Stochastic Differential Equations},
  booktitle = {International Conference on Learning Representations (ICLR)},
  year = {2021}
}

@article{Amblard2011mfBm,
  author  = {Amblard, Pierre-Olivier and Coeurjolly, Jean-Fran{\c{c}}ois},
  title   = {Identification of the Multivariate Fractional Brownian Motion},
  journal = {IEEE Transactions on Signal Processing},
  volume  = {59},
  number  = {11},
  pages   = {5152--5168},
  year    = {2011},
  doi     = {10.1109/TSP.2011.2162835},
  }

@article{SelaHurvich2009,
  author  = {Sela, Rebecca J. and Hurvich, Clifford M.},
  title   = {Computationally Efficient Methods for Two Multivariate
             Fractionally Integrated Models},
  journal = {Journal of Time Series Analysis},
  volume  = {30},
  number  = {6},
  pages   = {631--651},
  year    = {2009},
  doi     = {10.1111/j.1467-9892.2009.00631.x}
}

@article{Sonmez2018,
  author  = {S{\"o}nmez, Ercan},
  title   = {The Hausdorff dimension of multivariate operator-self-similar Gaussian random fields},
  journal = {Stochastic Processes and their Applications},
  volume  = {128},
  pages   = {426--444},
  year    = {2018},
  doi     = {10.1016/j.spa.2017.05.003}
}

@article{Bierme2007,
  author  = {Bierm{\'e}, Hermine and Meerschaert, Mark M. and Scheffler, Hans-Peter},
  title   = {Operator scaling stable random fields},
  journal = {Stochastic Processes and Their Applications},
  volume  = {117},
  number  = {3},
  pages   = {312--332},
  year    = {2007},
  doi     = {10.1016/j.spa.2006.07.004}
}

@article{Didier2011,
  author  = {Didier, Gustavo and Pipiras, Vladas},
  title   = {Integral representations and properties of operator fractional {Brownian} motions},
  journal = {Bernoulli},
  volume  = {17},
  number  = {1},
  pages   = {1--33},
  year    = {2011},
  doi     = {10.3150/10-BEJ259}
}

@article{Meraz2022,
  author  = {Meraz, M. and Alvarez-Ramirez, J. and Rodriguez, E.},
  title   = {Multivariate rescaled range analysis},
  journal = {Physica A: Statistical Mechanics and its Applications},
  volume  = {589},
  pages   = {126631},
  year    = {2022},
  doi     = {10.1016/j.physa.2021.126631}
}
\end{document}